\documentclass[aps,prl,
amsmath,
amssymb,showpacs,showkeys,reprint]{revtex4-1}

\usepackage{graphicx}
\usepackage{epstopdf}
\usepackage{upgreek}

\begin{document}
\preprint{APS/123-QED}


\title{A Matterwave Transistor Oscillator}


\author{Seth C. Caliga}
\author{Cameron J. E. Straatsma}
\author{Alex A. Zozulya}
\altaffiliation[Permanent address: ]{Department of Physics, Worcester Polytechnic Institute, Worcester MA 01609-2280}
\author{Dana Z. Anderson}
\affiliation{Department of Physics, University of Colorado, and JILA, University of Colorado and National Institute for Standards and Technology, Boulder CO 80309-0440}


\date{\today}

\begin{abstract}
An atomtronic transistor circuit is used to realize a driven matterwave oscillator. The transistor consists of \textquotedblleft{Source}\textquotedblright and \textquotedblleft{Drain}\textquotedblright regions separated by a narrow \textquotedblleft{Gate}\textquotedblright  well.  Quasi-steady-state behavior is determined from a thermodynamic model, which reveals two oscillation threshold regimes. One is due to the onset of Bose-Einstein condensation in the Gate well, the other is due to the appearance of a negative transresistance regime of the transistor.  The thresholds of oscillation are shown to be primarily dependent on the potential energy height difference between Gate-Drain and Gate-Source barriers.  The transistor potential is established with a combination of magnetic and optical fields using a compound glass and silicon substrate atom chip.  The onset of oscillation and the output matterwave are observed through in-trap imaging. Time-of-flight absorption imaging is used to determine the time dependence of the Source well thermal and chemical energies as well as to estimate the value of the closed-loop ohmic Gate resistance, which is negative and is observed to cause cooling of Source atoms.  
\end{abstract}
\maketitle
The classic electronic oscillator utilizes gain and feedback in conjunction with a power source to achieve sustained oscillation from frequency selective components~\cite{Armstrong:1914vy, Armstrong:1926wt,Meissner:1919wb,Meissner:1933ws}.  Here, we utilize those same heuristic elements to implement an \textit{atom} analog of a single-transistor electronic driven oscillator circuit.  When appropriately coupled to an antenna or a waveguide an electronic oscillator emits an electromagnetic wave, the energy of which is carried by photons.  In an analogous fashion, our atom oscillator emits a matterwave, the energy of which is carried by atoms. In contrast to a \textit{forced} oscillator such as a playground swing under the command of a child, a \textit{driven} oscillator incorporates a gain-induced dynamical instability --- an instability that is typically intended to drive the system away from a non-oscillating equilibrium state to an oscillating one. Familiar driven oscillators include magnetrons, masers, and lasers, as well as many types of semiconductor device based circuits~\cite{MacEVanValkenburg:2001vm}.  
\begin{figure}[floatfix]
\includegraphics[width=\columnwidth]{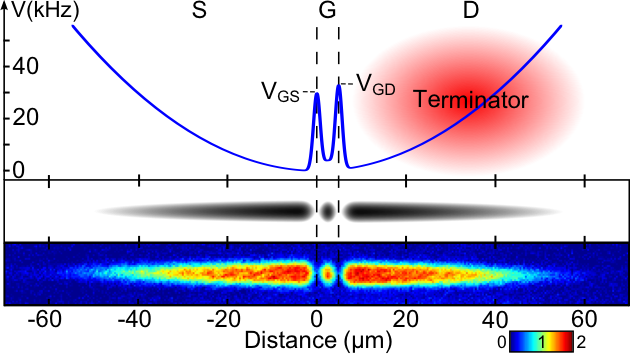}
\caption{\label{fig:Potential} The atomtronic transistor.  A hybrid magnetic and optical potential creates the overall confinement of the atomic ensemble as well as Source, Gate, and Drain wells labeled S, G, and D, respectively. The upper panel shows the loose magnetic confinement along with the $2.1~\upmu\text{m}$ full-width at $1/e$  blue-detuned optical potential barriers separated by $4.8~\upmu\text{m}$ used to create the Gate well. The terminator beam, illustrated in the Drain, out-couples atoms from the magnetic trap by optically pumping them into an untrapped $m_F$ state. The middle panel is a calculated potential energy density plot while the lower panel shows a false color in-trap absorption image of atoms occupying all three wells.  An optical density scale is shown below the horizontal axis.}  
\end{figure}

Gain in the atom-based system is provided by a triple-well potential atomtronic transistor~\cite{Stickney:2007ix,Pepino:2009jb,Seaman:2007kx}.  The top panel of Fig.~\ref{fig:Potential} shows the longitudinal profile of the potential energy surface that defines the transistor. Here, and throughout this work, energy is reported in units of Hz. Borrowing the nomenclature of field-effect transistors (FETs) we label the three regions of the atomtronic transistor as the Source, Gate, and Drain. The middle panel of Fig.~\ref{fig:Potential} is a calculated density plot of the potential with the outer boundary corresponding to 30 kHz. The bottom panel provides a false color in-trap absorption image of approximately $4.5\times10^{4}$  $^{87}{\rm{Rb}}$ atoms that have been loaded into the trap then subsequently cooled by RF forced evaporation to about $1~\upmu\text{K}$. The image reveals the three distinct regions of the transistor. The potential is created using an atom chip that produces a cigar-shaped magnetic trap.  Optical access to the trap is provided by an on-chip window. It allows a pair of blue-detuned ($760~\text{nm}$) optical barriers to be projected onto the magnetic trap, slicing it into the three regions~\cite{Salim:2013gb}.  The barriers are created using an acousto-optic modulator driven by a pair of independent radio-frequency generators, which allows the two barrier positions and heights to be adjusted independently.

We model the circuit beginning with the assumption that the Source and Gate wells have definite temperatures and chemical potentials denoted by $T_{\text{s}}$ and $T_{\text{g}}$ and $\mu_{\text{s}}$ and $\mu_{\text{g}}$, respectively. The barrier energies between the Source and the Gate ($V_{\text{GS}}$) and between the Gate and the Drain ($V_{\text{GD}}$) are much larger than the thermal or chemical energies of the particles in the well.  Thus coupling between the wells is via atoms energetic enough to traverse the barriers.  Building from the works of Luiten et al.  \cite{Luiten:1996ja} and Roos et al. \cite{Roos:2003tp} we take the atom currents to be given by
\begin{eqnarray}\label{currents}
   && I_{\text{sg}}=\gamma_{\text{s}} N_{\text{se}} exp[-(V_{\text{GS}}-\mu_{\text{s}})/T_{\text{s}}], \nonumber \\
   && I_{\text{gs}}=\gamma_{\text{g}} N_{\text{ge}} exp[-(V_{\text{GS}}-\mu_{\text{g}})/T_{\text{g}}], \nonumber \\
   && I_{\text{gd}}=\gamma_{\text{g}} N_{\text{ge}} exp[-(V_{\text{GD}}-\mu_{\text{g}})/T_{\text{g}}],
\end{eqnarray}
where $\gamma_{\text{s}}$ and $\gamma_{\text{g}}$ are effective collision rates, and $N_{\text{se}}$ and $N_{\text{ge}}$ are thermal atom numbers in the Source and Gate, respectively. Note that the subscript order indicates the direction of the current, e.g. $I_{\text{sg}}$ indicates current flowing from the Source to the Gate.  Particle number and energy conservation are expressed as
\begin{eqnarray}\label{Kirchoff}
  I_{\text{sg}}&&=I_{\text{gs}}+I_{\text{gd}}, \\
 I_{\text{sg}}&&(V_{\text{GS}}+\kappa_{\text{GS}}T_{\text{s}}) \nonumber \\
 \label{Energy}    &&= I_{\text{gs}}(V_{\text{GS}}+\kappa_{\text{GS}}T_{\text{g}})+I_{\text{gd}}(V_{\text{GD}}+\kappa_{\text{GD}}T_{\text{g}}),
\end{eqnarray}
where the $\kappa$'s, which indicate the average excess energy of atoms traversing the barriers, are of order unity \cite{Roos:2003tp}.  Since the Source contains a large number of atoms, its temperature and chemical potential will vary slowly in time compared to those of the Gate; therefore, we seek the dynamical quasi-equilibrium values of the Gate temperature and chemical potential in terms of those of the Source. It proves useful to define a temperature drop and barrier height difference normalized to the Source temperature:
\begin{eqnarray}\label{DeltaT}
  &&\tau  \equiv \left( {{T_{\text{s}}} - {T_{\text{g}}}} \right)/{T_{\text{s}}},  \\
&&v\equiv (V_{\text{GD}}-V_{\text{GS}})/T_{\text{s}}.
\end{eqnarray}
We will refer to $v$ as the \textquotedblleft{feedback}\textquotedblright parameter.  Eqs.~(\ref{Kirchoff}) and~(\ref{Energy}) can be solved in steady-state, and the temperature drop is determined from the following transcendental relationship:
\begin{figure}[floatfix]
\includegraphics[width=\columnwidth]{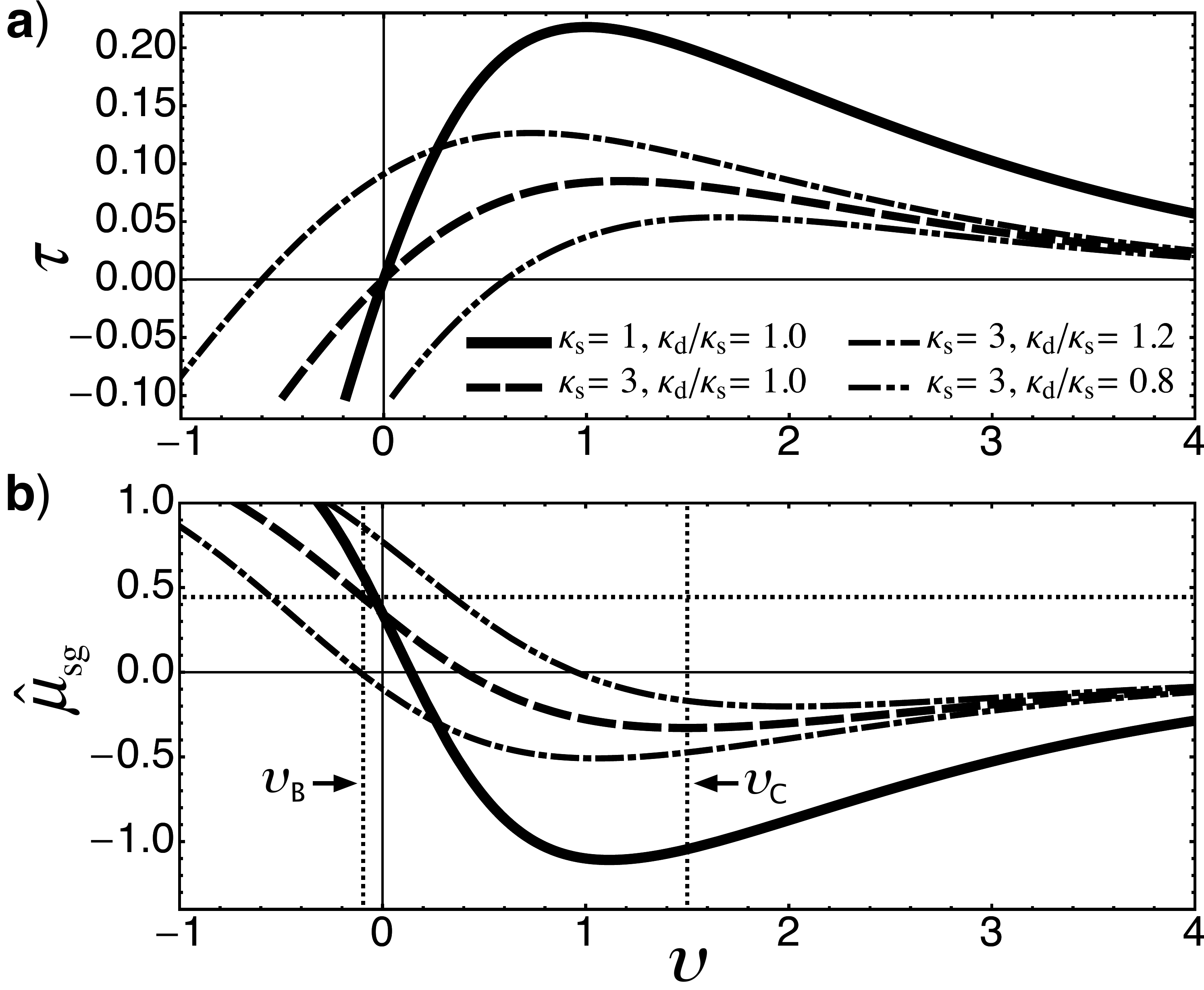}
\caption{\label{fig:Temperature_Drop} Dependence on feedback of a) the Source-Gate temperature drop and b) the Source-Gate potential drop.  The drops are calculated for various values of excess energy factors.  The potential drop is calculated using $\hat{\mu}_{\text{s}}=1$ and $\hat{V}_{\text{GS}}=5.5$, corresponding to nominal experimental values.  The thresholds for the onset of BEC ($v_{\text{B}}$) and negative transconductance ($v_{\text{C}}$) are shown for the $\kappa_{\text{GS}},\kappa_{\text{GD}}=3$ case. }
\end{figure}
\begin{equation}\label{DeltaTrans}
    \tau  = {e^{ - v/\left( {1 - \tau } \right)}}\frac{{v + \left( {{\kappa_{\text{GD}}} - {\kappa_{\text{GS}}}} \right)}}{{{\kappa_{\text{GS}}} + {\kappa_{\text{GD}}}{e^{ - v/\left( {1 - \tau } \right)}}}}.
\end{equation}
With the temperature drop in hand, the potential drop $\mu_{\text{sg}}\equiv\mu_{\text{s}}-\mu_{\text{g}}$ can also be calculated from the transcendental relationship
\begin{eqnarray}\label{PotentialDrop}
&&\hat{\mu }_{sg}=-\tau  \left(\hat{V}_{\text{GS}}-\hat{\mu }_{\text{s}}\right)+\left(1-\tau \right)\ln\left[\left(1-\tau\right)^{4}\left(1+e^{-\frac{v}{1-\tau }}\right)\right]  \nonumber \\
&&+(1-\tau )\ln \left[\frac{1-\tau+\frac{\zeta(2)}{\zeta(3)} \left(\hat{\mu }_{\text{s}}-\hat{\mu }_{\text{G0}}-\hat{\mu }_{sg}\right)}{1+\frac{\zeta(2)}{\zeta(3)} \left(\hat{\mu }_{\text{s}}-\hat{\mu }_{\text{G0}}\right)}\right],~~~
\end{eqnarray}
where the hat ($\hat{~}$)  indicates quantities normalized to the Source temperature, $\zeta(z)$ is the Riemann zeta function and $\mu_{\text{G0}}$ is the Gate ground state energy.  In deriving Eq.~(\ref{PotentialDrop}) we have used the fact that the thermal and potential drops are zero for $v=\infty$ along with reference \cite{Pethick:2002tn} to determine the ratio $\gamma_{\text{g}}N_{\text{ge}}/\gamma_{\text{s}}N_{\text{se}}$.  Fig.~\ref{fig:Temperature_Drop} shows the temperature and potential drops as a function of feedback using $\kappa_{\text{GS}}$, $\kappa_{\text{GD}}=1$ as a reference as well as the trend with varying combinations of excess energy factors. As suggested by the figure, there will generally exist a region of barrier difference for which the potential drop is negative, meaning the Source-Gate junction is \textit{reverse biased}.

In analogy with electronic circuits, we introduce an ohmic Gate resistance, $R_{\text{g}}\equiv\mu_{\text{sg}}/I_{\text{gd}}$, and the \textit{transresistance}
\begin{eqnarray}\label{tranresistance}
  r_{\text{g}}  \equiv  \frac{d\mu _{\text{sg}}}{dI_{\text{gd}}}=\frac{\partial\mu_{\text{sg}}}{\partial{T_{\text{s}}}}\left(\frac{\partial{I_{\text{gd}}}}{\partial{T_{\text{s}}}}\right)^{-1},
\end{eqnarray}
where the partial derivatives are evaluated at constant $\mu_{\text{s}}$. Both $R_{\text{g}}$ and $r_{\text{g}}$ can be negative. Negative resistance is a familiar concept in circuit theory, introduced  early in the literature on electronic oscillators ~\cite{Edson1953}. Power dissipation $P_{\text{d}}={I_{\text{s}}}^{2}R_{\text{g}}$ through a negative ohmic resistor indicates cooling rather than the usual heating.  Negative transresistance heralds impending instability: oscillation can be sustained if the negative transconductance is sufficiently large to compensate for the natural damping of a resonant circuit, and excessively large values can lead to multimode behavior as well as nonlinear phenomena such as harmonic generation and self-pulsing. 

Whereas electronic and optical oscillators excite modes of the electromagnetic field of a resonator, we are concerned with the excitation of matterwave modes of the Gate.  As the feedback is increased from $v=-\infty$ the model predicts two distinct thresholds, which are indicated in Fig.~\ref{fig:Temperature_Drop} for the $\kappa_{\text{GS}},\kappa_{\text{GD}}=3$ case.  The first, $v=v_{\text{B}}$, corresponds to the onset of Bose-Einstein condensation (BEC) in the Gate and is determined by the value of the feedback for which the Gate chemical potential is equal to the Gate's ground state energy, $\mu_{\text{g}}=\mu_{\text{G}0}$.  For sufficiently low Source temperatures, the transistor output will excite a single transverse mode of the Drain and the output will thus be spatially coherent.  Its temporal coherence will be dictated by the Gate temperature, which, interestingly, can be lower than if there were no Gate-Source barrier present.  

\begin{figure}[floatfix]
\includegraphics[width=\columnwidth]{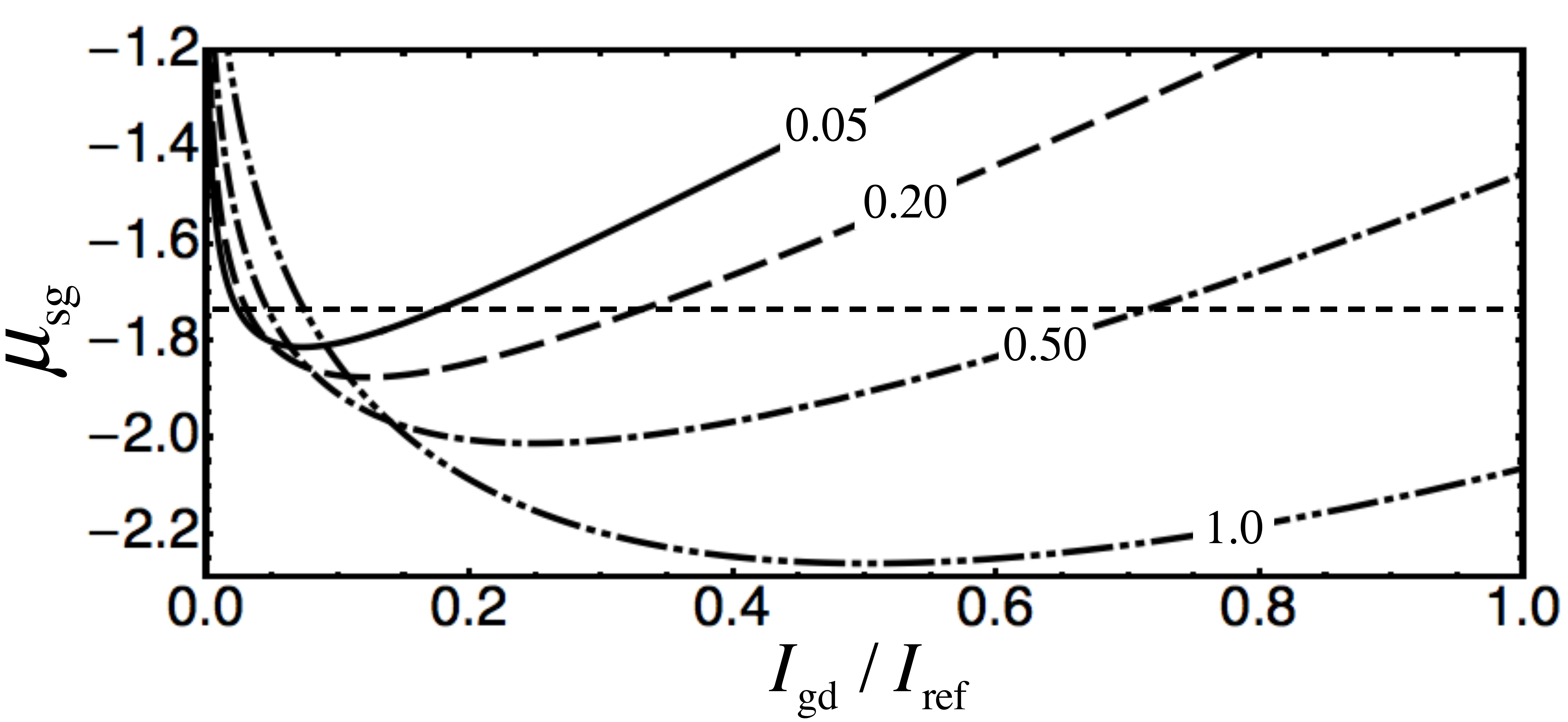}
\caption{\label{fig:MuICurves} Negative transresistance can be seen in this series of $\mu$-$I$ curves each labeled with its fractional value above threshold, $v_{\text{C}}$.  The horizontal dotted line indicates the equilibrium Source-Gate potential drop at the threshold.  The reference current $I_{ref}$ is chosen for plotting convenience. Plots are calculated using $V_{\text{GS}}=30$ kHz, $\mu_{\text{s}}=T_{\text{s}}=5.5$ kHz, $\kappa_{\text{GS}}, \kappa_{\text{GD}}=3$.}
\end{figure}

With further increase of the feedback, a second threshold, $v=v_{\text{C}}$, arises as the transresistance becomes negative, and the equilibrium solutions depicted in Fig.~\ref{fig:Temperature_Drop} can become unstable.  To the extent that the Gate well is harmonic one should expect a strong response of the dipole mode of oscillation at the longitudinal trap frequency, since this mode has the lowest resonant frequency and  its damping is small ~\cite{Griffin:2009wva}. More generally one might expect multimode oscillation. The dynamical analysis of the circuit's behavior is outside the scope of our model, yet is of considerable interest since it is conceivable that appropriate circuit design can give rise to a temporally coherent output.  Certainly the analogy with electronic oscillators suggests this possibility.    Fig.~\ref{fig:MuICurves} provides some insight into this interesting regime of circuit operation using a series of $\mu$-$I$ curves above threshold, the slopes of which indicate the transresistance.  Worth noting is the fact that as feedback is increased, the point of zero transresistance moves to increased reverse bias and output current.  
\begin{figure}[floatfix]
\includegraphics[width=\columnwidth]{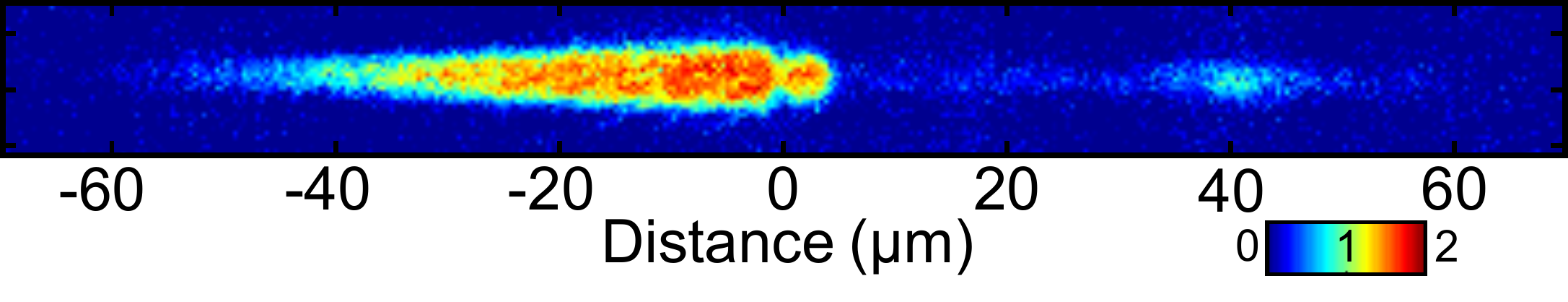}
\caption{\label{fig:Hold} Five shot average of in-trap absorption images showing the atomtronic oscillator after $30~\text{ms}$ operation time for above-threshold barrier height difference.  The initially empty Gate well shows high optical density while emitted atoms appear in the Drain, with the peak atom number near the classical turning point at $40~\upmu\text{m}$.}
\end{figure}

The experimental work has been carried out using the trap shown in Fig.~\ref{fig:Potential}. The longitudinal trap frequency of the Gate is $\nu_{\text{G}}\simeq850~\text{Hz}$  while its total ground-state energy is shifted by optical and magnetic biases to $\mu_{\text{G}0}=3.0~\text{kHz}$.  The longitudinal frequency of the magnetic trap without the barriers is $67~\text{Hz}$, and therefore with the barriers the longitudinal frequency is approximately $\nu_{\text{S}}=134~\text{Hz}$ for the Source and $\nu_{\text{D}}=155~\text{Hz}$ for the Drain.  The transverse frequencies of all three wells are approximately  $\nu_{\perp}=1.7~\text{kHz}$. The oscillator is prepared by creating a BEC in the Source well with predetermined temperature and chemical potential using RF forced evaporation~\cite{Salim:2011bq}. The barriers are kept sufficiently high during preparation so that the Gate and Drain wells remain empty. During operation the Drain region of the transistor is illuminated, as depicted in Fig.~\ref{fig:Potential}, with laser light tuned to atomic resonance.  This ``terminator'' beam causes atoms to leave the trap; thus, atoms entering the Drain from the Gate are effectively coupled to the matterwave impedance of the vacuum \cite{Mokhov:2008cv}.

Time $t=0$ is defined as the time that the barriers are lowered from their BEC preparation heights to their operating configuration.  A measurement is carried out by allowing the circuit to evolve for a time $\Delta{t}$ and then imaging the atoms.  Data is taken either from time-of-flight absorption images using standard techniques~\cite{Anderson:1995vb,Pappa:2011ig}, or from in-trap absorption images.  Time-of-flight measurements provide information about the Source temperature and chemical potential.  For in-trap imaging the terminator beam is extinguished $5~\text{ms}$ before acquiring an image, which is somewhat less than half of the oscillator period in the barrier-free potential. Because atoms propagate slowly at the top of their climb the probability density is maximum near the classical turning point, where the potential energy of the trap is equal to the atom's kinetic energy.  

Fig.~\ref{fig:Hold} shows an in-trap absorption image taken after $30~\text{ms}$ of hold time with $V_{\text{GS}}=30~\text{kHz}$ and $V_{\text{GD}}=33~\text{kHz}$. One can observe the atoms emitted from the Gate at a position of about $40~\upmu\text{m}$, corresponding to an energy of $33~\text{kHz}$.  One can also see that atoms have settled into the Gate.  Though it would be ideal to have a quantitative measure of the Gate atom number, our system has been optimized for projection rather than imaging~\cite{Reinaudi:2007ut,Pappa:2011ig}.  Nevertheless, note that the Gate peak optical density is comparable to that of the source.  Since the Gate has a large positive bias compared with $\mu_{\text{s}}$, one can conclude that $\mu_{\text{sg}}$ is strongly reversed biased during operation, as suggested in Fig.~\ref{fig:MuICurves}.  This is discussed further below.
\begin{figure}
\includegraphics[width=\columnwidth]{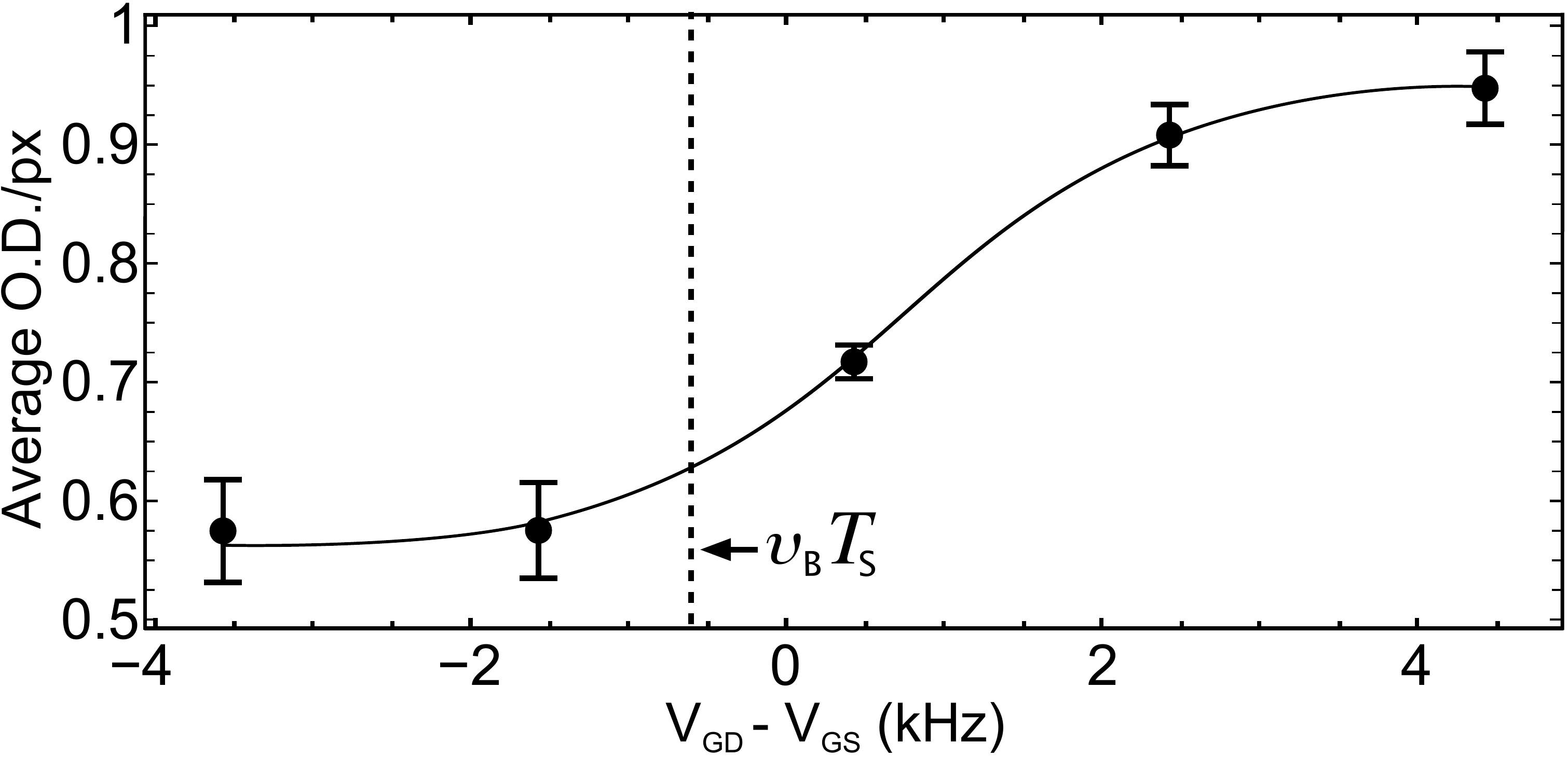}
\caption{\label{fig:Threshold} In-trap measured average Gate optical density (O.D.) observed after $30~\text{ms}$ circuit operation time as a function of barrier height difference.  Error bars are the standard error of the mean for three measurements taken for each data point. The initial Source temperature and chemical potential are 9.8 kHz and 4.4 kHz respectively. The vertical dotted line indicates the calculated BEC threshold. The solid curve is meant to serve as a visual aid only.}  
\end{figure}
\begin{figure}
\includegraphics[width=8.6cm]{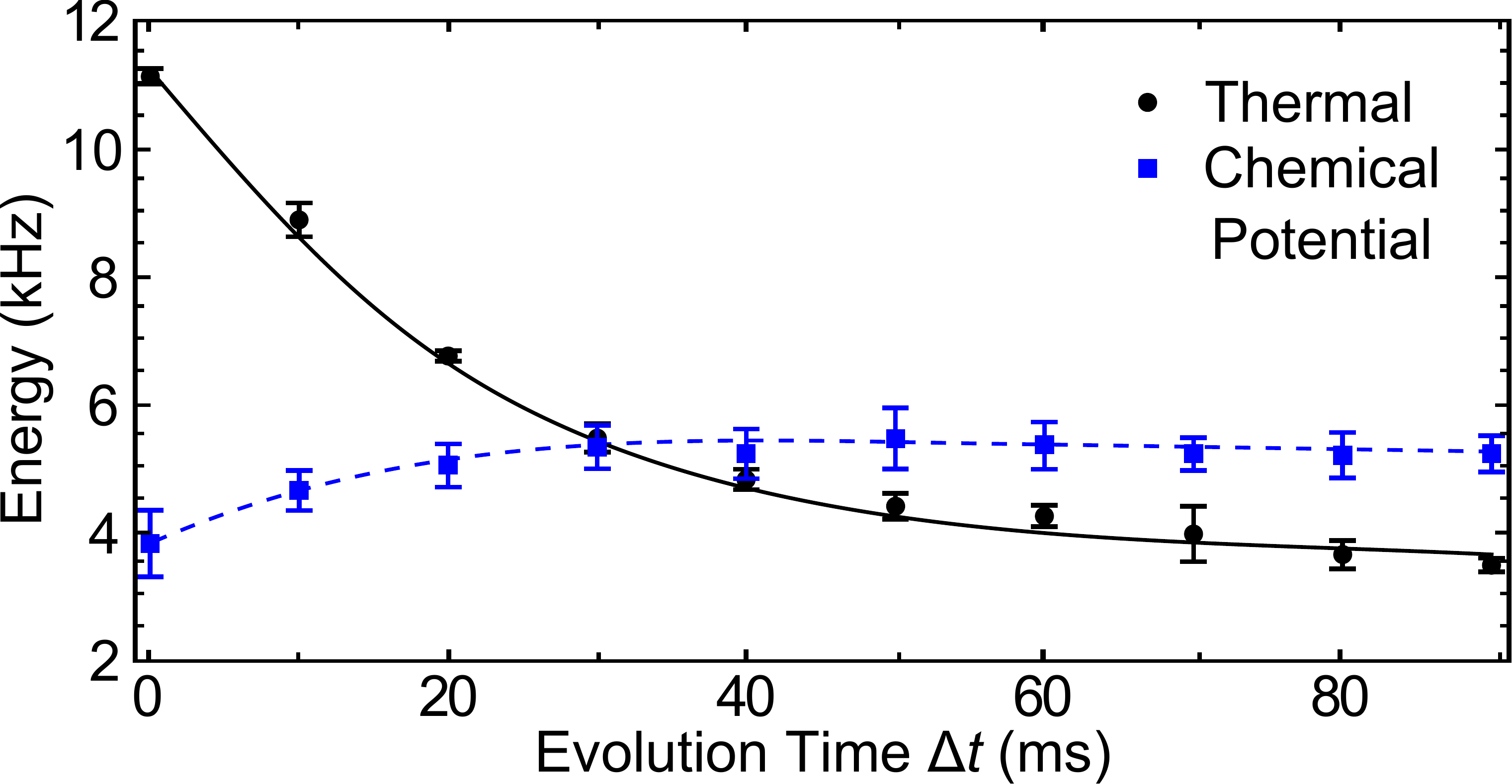}
\caption{\label{fig:WellEnergies} Source well populations and energies during circuit oscillation derived from time-of-flight imaging. Error bars are the standard error of the mean for three measurements taken for each data point.The curves are meant to serve as a visual aid only.}  
\end{figure}
Fig.~\ref{fig:Threshold} shows the measured optical density of the Gate as a function of the barrier height difference, $V_{\text{GD}}-V_{\text{GS}}$, where $V_{\text{GD}}$ is changed while holding $V_{\text{GS}}=30~\text{kHz}$.  The plot reveals an S-shaped turn-on character, which is typical of driven oscillators.   

Our model treats the Source well essentially as a ``battery" with fixed temperature and chemical potential.  In fact, these quantities evolve as the Source loses atoms and also undergoes cooling due to the Gate feedback current, $I_{\text{gs}}$. Fig.~\ref{fig:WellEnergies} shows the time dependence of the thermal and chemical energies of atoms in the Source determined from time-of-flight measurements for $V_{\text{GS}}=30~\text{kHz}$ and $V_{\text{GD}}=33~\text{kHz}$.   It is informative to compute the oscillator parameters at  $t=30$ ms, which is the oscillation time corresponding to Fig.~\ref{fig:Hold}. For converting between normalized and unnormalized quantities we will use the 30 ms Source temperature $T_{\text{s}}=5.5$~kHz from Fig.~\ref{fig:WellEnergies}.  Thus the feedback parameter is about $v=0.58$.  The Source chemical potential $\mu_{\text{s}}\simeq5.5$~kHz. The calculations of Roos et al. estimate that $\kappa_{\text{GS}}\simeq\kappa_{\text{GD}}\equiv\kappa\simeq2.9$ for our trap configuration~\cite{Roos:2003tp}; although those results apply to a different density regime, we will take $\kappa=3$ for calculations using Eqns.~(\ref{DeltaTrans}) and (\ref{PotentialDrop}).  Using the data of Fig.~\ref{fig:WellEnergies} we find that the fractional temperature drop of the Gate relative to the Source is about 7\% after a $30~\text{ms}$ oscillation time.  The potential drop calculated from Eq.~(\ref{PotentialDrop}) is $\mu_{\text{sg}}=-0.51$ kHz which corresponds to a gate potential of $\mu_{\text{g}}=6.0\text{~kHz}$.  Subtracting the gate bias indicates the optical density of the gate corresponds to a chemical potential of $(\mu_{\text{g}}-\mu_{\text{G}0})=3.0$ kHz which is below the Source potential by 2.5 kHz. This is consistent with Fig.~\ref{fig:Hold} in which the Gate and Source are seen to have similar peak optical densities. 

Fig.~\ref{fig:Temperature_Drop} indicates $v_{\text{B}}=-0.11$, corresponding to an un-normalized barrier height difference of $(V_{\text{GD}}-V_{\text{GS}})=-600$~Hz, which is indicated on Fig.~\ref{fig:Threshold}. In-trap images of the gate do not provide a definitive threshold for the onset of BEC in the gate; the predicted -600 Hz lies well below the middle of the \textquotedblleft{S}\textquotedblright of the experimental turn-on curve, so it is not inconsistent with the observations.  The maximum reverse bias is predicted to occur at $(V_{\text{GD}}-V_{\text{GS}})=8.3$ kHz, while the observed maximum occurs at 4.5 kHz. Evidently the optical density is saturated; whether this is due to a physical mechanism or to limitations of the high NA, low depth of focus imaging system has yet to be determined.

Time-of-flight measurements indicate a current of  $I_{\text{gd}}\simeq120~\text{kHz}$ at $t=30~\text{ms}$. This gives us the resistance in dimensionless units for energies reported in units of Hz: $R_{\text{g}}(t=30~\text{ms})=-4.2\times{10}^{-6}$. Cooling of the Source atoms is a consequence of this negative resistance, which can be understood by calculating the power dissipated: $P_{\text{d}}=I_{\text{gd}}^2R_{\text{g}}\simeq-6.61\times{10^4}$ Hz$^2$.  The effects of negative power dissipation can be seen in Fig.~\ref{fig:WellEnergies} not only as a decrease in Source temperature but also as an increase in chemical potential. 

Our current experiments have focused on the BEC oscillator regime in which the feedback is kept below the negative transresistance threshold.  While it is convenient to consider circuits in terms of bulk rather than distributed elements, the output matterwave wavelength is on the order of $\lambda_{\text{D}}\simeq{250}$ nm. Our circuit is therefore more akin to an electronic microwave oscillator than to an audio one.  It is not surprising, therefore, that instabilities would cause excitations of collective modes in the gate; by the same token, circuit behavior can be impacted by the impedance on both the Source side and the Drain side of the gate~\cite{Mokhov:2008cv}.   Much could be learned from experiments designed to measure the spectrum of the output wave, and considerable theoretical work is needed to better understand the dynamical nature of the atomtronics oscillator in the large-feedback regime.

This work is supported by AFOSR, NSF, and DARPA. The work of S. Caliga, A. Zozulya and D.Z. Anderson was partially supported by the Charles Stark Draper Laboratories. 

\bibliography{Oscillator_21020818}

\end{document}